\documentclass{aa}
\usepackage{graphicx}
\usepackage{txfonts}
\usepackage{textcomp}
\usepackage{amsmath}

\begin{document}

\title{Spatial distribution of exoplanet candidates based on Kepler and $Gaia$
data} 
\titlerunning{Spatial distribution of exoplanet candidates} 
\author{A.~Maliuk \inst{1}\and J.~Budaj\inst{1}}
\institute{Astronomical Institute, Slovak Academy of Sciences,
05960 Tatransk\'{a} Lomnica, Slovak Republic\\
\email{amaliuk@ta3.sk, budaj@ta3.sk}}  
\date{Received ???? ??, ????; accepted ???? ??, ????}

\abstract 
{ Surveying the spatial distribution of exoplanets in the Galaxy is important 
for improving our understanding of planet formation and evolution.}
{We aim to determine the
spatial gradients of exoplanet occurrence in the Solar 
neighbourhood and in the vicinity of open clusters.}
{We combined Kepler and $Gaia$ DR2 data  for this purpose, splitting the volume sampled by the Kepler mission into certain
spatial bins. We determined  an uncorrected and bias-corrected
exoplanet frequency and metallicity for each bin.
}
{There is a clear drop in the  uncorrected exoplanet frequency with 
distance for F-type stars (mainly for smaller planets),
a decline with increasing distance along the Galactic longitude 
$l=90^{\circ}$,
and a drop with  height above the Galactic plane.
We find that the metallicity behaviour cannot be the reason 
for the drop of the exoplanet frequency around F stars with 
increasing distance. This might have only contributed to the drop in 
 uncorrected exoplanet frequency with the height above 
the Galactic plane.
We argue that the above-mentioned gradients of  uncorrected
exoplanet frequency are a manifestation of a single bias of undetected 
smaller planets around fainter stars.
When we correct for observational biases, most of these gradients
in exoplanet frequency become statistically insignificant.
 Only a slight decline of the planet occurrence with distance
for F stars remains significant at the $3 \sigma$ level.
Apart from that, the spatial distribution of exoplanets in the Kepler
field of view is compatible with a homogeneous one. 
At the same time, we do not find a significant change in 
the exoplanet frequency with increasing distance from open clusters.
In terms of byproducts, we identified six exoplanet host star candidates 
that are members of open clusters. 
Four of them are in the NGC 6811
(KIC 9655005, KIC 9533489, Kepler-66, Kepler-67)
and two belong to NGC 6866 (KIC 8396288, KIC 8331612).
Two out of the six had already been known to be cluster members.
}
{}

\maketitle

\section{Introduction and motivation}

The stellar environment within our Galaxy is far from homogeneous
and isotropic.
The Galaxy has a spiral structure and the disc undergoes 
large-scale perturbations caused by the spiral arms.
Another important source of inhomogeneity is in the form of shock waves generated 
by supernova explosions. These events enrich the interstellar medium with 
heavy elements and work as a trigger for star and planet formation.
Even within the disk, there are stars of different ages, populations,
and metallicities. 
For field stars, the age-metallicity relation is nearly flat up to 
8 Gyr with a clear drop in the metallicity for older stars.
Metallicity decreases with the Galactocentric radius and height above
the Galactic plane \citep{bergemann14,duong18}.
The nitrogen and oxygen abundances of Galactic HII regions were also
found to decrease with the Galactocentric radius 
\citep{esteban17,esteban18}.
Open clusters constitute 'islands' of stars with homogeneous
ages and metallicity. They differ from neighbourhood field stars
by the enhanced spatial density of their stars.
The metallicity of open clusters decreases with the distance from 
the Galaxy centre and increases with the age of 
the cluster \citep{netopil16,jacobson16}.
Such inherent inhomogeneity of the environment may have an impact
on planet formation and occurrence.
For example, the frequency of the exoplanets occurrence
depends on metallicity. Short-period gas giants
(hot Jupiters and warm sub-Neptunes)
are more likely to be found around metal-rich stars while smaller planets 
are found around stars with a wide range of metallicities \citep{fischer05,
buchhave12,narang2018,petigura2018} .
Planets may migrate over the course of their formation and their evolution,
which impacts the chances of their detection significantly.
Hot Jupiters were most probably born beyond the snow line and 
migrated inward, affecting all the inner planets. 

Unfortunately, we know very little about young exoplanets. 
\cite{eyken2012} found a transiting exoplanet candidate orbiting 
a T Tau star in the Orion-OB1a/25-Ori region. 
\cite{meibom13} discovered two mini-Neptunes (Kepler-66, Kepler-67) 
in the 1 Gyr cluster NGC 6811 which is in the Kepler field of view. 
They concluded that the frequency of planets in this cluster 
is approximately equal to the field one. 
\cite{kurtis2018} identified a sub-Neptune exoplanet 
transiting a solar twin EPIC 219800881(K2-231) in the Ruprecht 147 
stellar cluster. This indicates an exoplanet frequency
of the same order of magnitude as in NGC 6811.
\cite{quinn12} detected two hot Jupiters in the Praesepe cluster
and estimated a lower limit of 3.8+5.0-2.4\% on the hot Jupiter 
frequency in this metal-rich open cluster. Given the known age of 
the cluster, this also demonstrates that
giant planet migration occurred within 600 Myr after the formation.
\cite{libralato2016} presents the sample of seven exoplanet candidates
discovered in the Praesepe field. Two of them, K2-95 and EPIC 211913977,
are members of the cluster.
\cite{mann2017} found seven transiting planet candidates in Praesepe
cluster from the K2 light curves (K2-100b, K2-101b, K2-102b, K2-103b,
K2-104b, EPIC 211901114b, K2-95b). Six of them were confirmed
to be real planets, with the last one requiring more data. 
K2-95b was also studied in \cite{pepper2018}. 
\cite{rizzuto2017} studied nine known transiting exoplanets in the 
clusters (Hyades, Upper Scorpius, Praesepe, Pleiades) and also
identified one new transiting planet candidate orbiting 
a potential Pleiades member. 
The lack of detected multiple systems in the young 
clusters is consistent with the expected frequency from
the original Kepler sample within our detection limits.
\cite{rizzuto2019} addressed the question of planet occurrence 
in the young clusters observed by the K2 mission.
Initial results indicate that planets around 650-750 Myr M-dwarfs
have inflated radii but a similar frequency of occurrence compared 
to their older counterparts.
However, the 125 Myr old Pleiades has a lower occurrence rate of 
short period planets.
In Praesepe, \cite{rizzuto2018} also discovered 
 a two-planet system of K2-264. Both planets are likely 
mini-Neptunes.
K2-264 is one of two multiple-planet systems found in the open clusters.
The other is K2-136, a triple transiting-planet system in the Hyades cluster
\citep{mann2018, livingston2018}. 
K2-136 system includes an Earth-sized planet, a mini-Neptune, and 
a super-Earth orbiting a K-dwarf.
\cite{gaidos2016} describes a 'super-Earth-size'  planet transiting
an early K dwarf star observed by the K2 mission. The host 
star, EPIC 210363145, was identified as a member of the Pleiades cluster, 
but a more detailed analysis of the star's properties did not confirm
its cluster membership.
\cite{vanderburg2018} reported the discovery of a long-period transiting
exoplanet candidate with the mass of about $6.5 M_{\oplus}$,
called HD 283869b, orbiting another K-dwarf in the Hyades cluster.

The YETI (Young Exoplanet Transit Initiative) project conducts searches for
transiting exoplanets in a number of young 
open clusters.  The detection rate is lower than expected,
which may be due to an intrinsic stellar variability or the true paucity 
of such exoplanets \citep{neuhauser11,errmann14,garai16,fritzewski16}.
The theoretical study of \cite {2001MNRAS.322..859B} indicates that
while planetary formation is heavily suppressed in the crowded 
environment of the globular clusters, less crowded systems such as 
open clusters should have a reduced effect on any planetary 
system. \cite{fuji2018} also explored the survival rates of planets 
against stellar encounters in open clusters by performing a series 
of N-body simulations of high-density and low-density open clusters, 
along with open clusters that grow via mergers of sub-clusters, 
and embedded clusters. 
They found that less than 1.5 \% of close-in planets within 1 AU and 
at most 7\% of planets within 1-10 AU from the star are ejected by 
stellar encounters in clustered environments.
The ejection rate of planets at 10-100 AU around FGKM-type 
stars reaches a few tens of percent.

Another piece of evidence to demonstrate that  planet formation is affected by 
the presence of a more distant stellar companion comes from 
the study of binary stars.
\cite{wang14, wang14b} found that the circumstellar planet 
occurrence in such systems is significantly lower
than in single stars, indicating that the planet formation is 
significantly suppressed in this case. 
In the end, the question of the spatial distribution of exoplanets 
and their host stars or, more precisely, the local frequency of their 
occurrence throughout the Galaxy presents an unsolved problem 
intimately linked to the planet's formation and evolution.

The Kepler mission provides the most complete and homogeneous sample of 
exoplanets and their host stars to date \citep{borucki10}.
On the other hand, the  recent second data release of $Gaia$  
\citep{2016A&A...595A...1G, 2018A&A...616A...1G} provides the most 
precise distances to the stars. Together, the Kepler and $Gaia$ data 
provide the best
information about the spatial distribution of exoplanets available at present.
The goal of this study is not to provide an absolute estimate
of exoplanet occurrence; rather, we aim to search for relative variations in the planet 
occurrence in the space on 
(a) longer scales spanning hundreds to thousands of parsecs or 
(b) shorter scales of tens of parsecs in the vicinity of open clusters.
As a by product of this analysis, we
identify exoplanet candidates which are members
of the open clusters in the Kepler field of view.

\section{Stellar sample}

We start with the list of the Kepler target stars (KSPC DR 25) from 
\cite {Mathur17}, which counts about 190,000 stars.
The positions of all these stars were cross-matched with the $Gaia$
DR2 positions in \cite {2018arXiv180500231B}.
They used the X-match service of the Centre de Donn\'ees astronomiques 
de Strasbourg (CDS) and applied the following criteria
to match the stars:
the difference in the position smaller than 1.5 arcseconds;
and the difference in the magnitudes smaller than two magnitudes.
For stars with multiple matches that satisfied these
criteria, the authors decided to keep those with the smallest angular 
separations. Apart from that, the following stars were removed from 
the sample:
stars with poorly determined parallaxes ($\sigma_{\pi}/{\pi} > 0.2$), 
stars with low effective temperatures ($T_{\rm eff} < 3000 K$),
stars with either extremely low gravity ($\log g < 0.1$) (in CGS units),
or a low-quality Two Micron All-Sky Survey \citep{cutri2003} photometry
 (lower than 'AAA').
Following this procedure, the sample contained $177,911$ Kepler target stars
and $3084$  Kepler host star candidates with $4044$ exoplanet candidates.

In the next step, we excluded giant stars from the sample.
We put the following limitations on stellar radius from 
\cite {Fulton17} :
$R_{star}/R_{\odot} < 10^{0.00025(T_{\rm eff}-5500)+0.20}$. 
Using this criterion, we rejected $57,743$ giants of all types.
The final list contains 120 168 Kepler target stars, including 
2562 Kepler host star candidates with 3441 exoplanet candidates.
Kepler target stars may have different spectral types.
Most of them are F, G, and K star.
These stars each have a different brightness and are seen up to different
distances, which might cause biases in our analysis.
That is why we further broke the stellar sample up into three categories:
F stars with $6000$K$ \le T_{\rm eff} \le 7500$K, 
G stars with $5200$K$ \le T_{\rm eff} \le 6000$K, and 
K stars with $3700$K$ \le T_{\rm eff} \le 5200$K. 
There are 35075 K-stars, 64525 G-stars, 17750 F-stars, and 2818 other 
types of stars in the final sample.

Apart from the star and planet sample, we used other stellar properties
such as the metallicities in the form of [Fe/H] from
the Kepler stellar properties catalogue KSPC DR 25 and
coordinates, effective temperatures, parallaxes, and $G$-band 
magnitudes from $Gaia$ DR 2 
\citep{2016A&A...595A...1G,2018A&A...616A...1G}.
We note that $Gaia$ DR2 parallaxes are affected by a zero-point offset 
\citep{arenou2018,riess2018,zinn2018,khan2019}. This offset 
was taken into account by adding $+0.029$ mas 
(global value of zero-point) to all parallaxes before 
they were converted to distances \citep{lindegren2018}.

\section{ Spatial gradients of the exoplanet frequency}
\subsection{Uncorrected exoplanet frequency behaviour}

\begin{figure}
\center{\includegraphics[width=\linewidth]{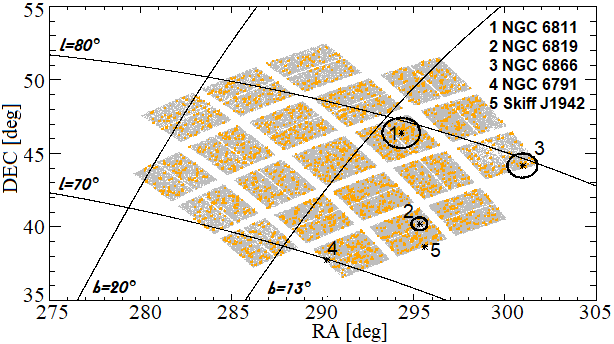}}
\caption{Location of exoplanet host stars (orange dots),
Kepler target stars (grey dots), and open clusters (asterisks) 
in the Kepler field of view.
Circles around some open clusters indicate the size of the inner
cylinders used for statistics; see  Sect.\ref{sec5} for more details.}
\label{figure:planets}
\end{figure}

To study the spatial distribution of exoplanets, we divided
the space into a number of smaller 3D segments according to
the right ascension $\alpha$, declination $\delta$, and distance
from the Sun $r$.
The Kepler field of view (see Fig. \ref{figure:planets}) is 
composed of 21 fields (associated with individual chips) with small 
gaps in between. That is why we created 21 spatial beams corresponding
to these fields.
Consequently, we split each beam into five segments according to
the distance.
In this way, we obtained 21$\times$5=105 spatial bins.
 In each bin, we calculated the ratio of the number of exoplanet 
candidates and Kepler target stars which we will call the uncorrected 
exoplanet frequency. We would like to point out that this frequency is not corrected for 
observing and completeness biases (see Sect. \ref{sec31}) and, thus, 
it is not a real exoplanet frequency, but a relative quantity 
proportional to it. We only use it as a guide to search for patterns 
that are worthy of more attention.
Then we assigned $(r,\alpha, \delta)$ coordinates to each bin
such that they were simply the centre of the bin.
We explored the gradients in different coordinate systems that is why
we assigned to the center of each bin also the $(r_g, z)$ coordinates
of the Galactocentric system and $(x, y, z)$ coordinates of 
the Cartesian system, where
$r_g$ is the projection of Galactocentric radius on the galactic plane
and $z$ is the distance from the galactic plane.
The $x$-axis of the Cartesian coordinate system  
is directed towards the centre of the Galaxy, $y$-axis is along 
the Galactic longitude 90\textdegree and $z$-axis is the same as above.
Then we approximated the  uncorrected exoplanet frequency with 
the following linear functions:
\begin {equation}
f(r,\alpha, \delta)=k_r r+k_{\alpha} \alpha+k_{\delta} \delta+k_0. 
\label{equation:linear}
\end {equation}
\begin {equation}
f(r_g,z)=k_{rg} r_g + k_z z + k_0 .
\label {equation:linear_g}
\end {equation}
\begin {equation}
f(x,y,z)=k_x x + k_y y + k_z z +k_0.
\label {equation:linear_xyz}
\end {equation}

\begin{figure}
\center{\includegraphics[width=\linewidth]{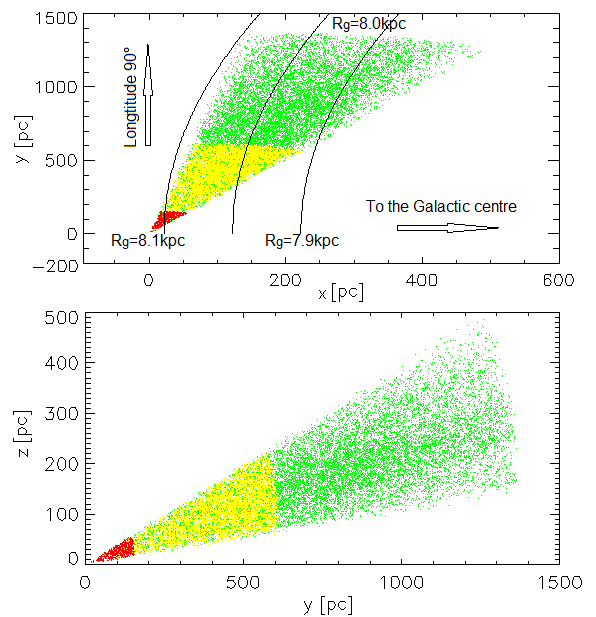}}
\caption{Location of the Kepler target stars in the Galactic disc 
in ($x, y$) plane (upper panel) and ($y, z$) plane (lower panel).
F, G, and K stars are highlighted with 
green, yellow, and red dots, respectively.
The X-axis points to the Galactic centre.
The direction of the Y-axis corresponds to the Galactic longitude of 
90$^\circ$. The three curves of constant Galactocentric radius are also 
plotted.}
\label{figure:position}
\end{figure}

When fitting equations 1-3, we assumed a Gaussian likelihood and used 
standard linear model fitting techniques for reporting uncertainties in
the parameters.  We adopted bins with no counts of a zero value during
the fit.  Since we used the results of this fit as an indicator of 
which relationships to explore with a higher fidelity model, in 
Sect. \ref{sec31}, we do not explore whether a more sophisticated model 
fitting (e.g. Poisson likelihood, upper limits, etc.) would 
significantly alter the fit model parameters.
Because F, G, and K stars have different brightness and are seen up
to different distances, we analysed them separately.
To create a volume-limited sample, we assumed some initial threshold 
visual $G$-band magnitude of 
16 mag and calculated the distances 
corresponding to typical F, G, and K main sequence stars.
The location of stars of different spectral class satisfying this
criterion is displayed in Fig.~\ref {figure:position} in a Cartesian
system aligned with the galactic coordinates.

The 16-magnitude threshold would correspond to a maximum distance of
250, 1000, and 2190 pc for K, G, and F stars, respectively.
Then we split this maximum distance into five equal bins as mentioned 
above. Apart from the brightness of the star, the planet detection 
efficiency depends on the transit depth and, hence, on the planet radius.
Planet occurrence itself may depend on the planet size.
That is why we also split the sample of exoplanets candidates
into the following intervals according to their radius:
$R_{planet} \ge 0.75R_{\oplus}$ , which covers most of the planets;
$0.75R_{\oplus} < R_{planet} \le 1.75R_{\oplus}$ , which covers
Earth-like planets and super-Earths;
$1.75R_{\oplus} < R_{planet} \le 3.0R_{\oplus}$ 
, which covers sub-Neptunes; and 
$R_{planet}> 3.0R_{\oplus}$ , which corresponds mostly 
to hot Jupiters. Then we fit for ($k_r$, $k_{\alpha}$,
$k_{\delta}$, $k_0$), ($k_{rg}$, $k_z$, $k_0$), and 
($k_x$, $k_y$, $k_z$, $k_0$) coefficients.
Apart from $k_0$, they correspond to the spatial gradients of 
the  uncorrected exoplanet frequency in different coordinate 
systems. The results are listed in 
Tables ~\ref{tab:table2}, ~\ref{tab:table3}, and ~\ref{tab:table12}. 

As can be seen in the tables, the exoplanet frequency gradients
along the $\alpha$ and $\delta$ coordinates are not statistically
significant. They are compatible with zero 
(within 2$\sigma$ errors).
\footnote{The $k_{\alpha}$ gradient of super-Earths around 
F stars 
might be just above the 2$\sigma$ limit.}
Nevertheless, we find a statistically significant negative $k_{r}$ value 
of the gradient in the distance $r$ for all the planets around F stars
except those with $R_{planet} \ge 3.0R_{\oplus}$.
We do not find any significant gradient of the exoplanet frequency 
with the Galactocentric radius. However, again,
for all the planets around the F stars 
(except those with $R_{planet} \ge 3.0R_{\oplus}$) we find 
statistically significant negative values of $k_z$, 
the gradient in the height 
above the Galactic plane. 
At the same time, we find significant negative gradients
along the $y$ axis for the same stars (F stars ). 
 Since we are observing the significant trends based on 
the uncorrected planet counts,we explore in 
Sect. \ref{sec31} whether these trends are intrinsic to the Galaxy or
whether the trends have been injected by detection biases.

\subsection{Discussion, bias correction, interpretation, 
and disentangling}
\label{sec31}

The transit method of exoplanet detection suffers from heavy biases.
Its efficiency depends on the planet-to-star radius ratio,
orbital period, eccentricity, inclination, and stellar brightness. 
 Such biases in the Kepler data were included in recent studies
by \cite{mulders18,sluijs18,zhu2018,petigura2018} 
or \cite{kipping2016}. Given that our sample is homogeneous. 
most of these biases
will be the same over the whole field of view, except for 
the distance bias.
Distance bias reduces the number of small planets detected with 
the distance from the observer since it is more difficult
to find small planets around fainter stars.
We notice that F stars are the ones that sample the largest distances.
It is also notable that the $k_r$ values gradually drop 
and become more significant with decreasing the planet radius, indicating that this trend may be due to an above-mentioned
observational distance bias.

Apart from the above remarks, we note a strong correlation
between the distance $r$ and the $y$-coordinate, as well as 
a correlation between the $y$ and $z$-coordinate for the stars
in the Kepler field of view (see Fig.\ref{figure:position}).
Consequently, any gradient along the distance might reflect mainly
onto a gradient along the $y$ coordinate of the Cartesian system
or along the $z$ coordinate of the Galactocentric system.
This is precisely what we found and that is why we need to correct
for such a bias.
We followed \cite{burke15}, who suggested a method 
of deriving an exoplanet occurrence which takes into account all 
aforementioned biases. Due to its numerical efficiency, we
adopted the \cite{burke15} model for the Kepler pipeline completeness 
rather than the more accurate and recent model from \cite{burke2017}.
 The \cite{burke15} model describes the algorithm for calculation of
probability that the transit event will be detected:
\begin {equation}
P_K(R,P)=\frac{1}{b^a \Gamma(a)}\int_{0}^{x}t^{a-1}e^{-t/b}dt
\label{equation:detect}
,\end {equation}
where $P_K$ is the  signal recoverability of the $Kepler$ pipeline
for planet with  
given radius $R$ and orbital period $P$. The best-fit coefficients 
to the sensitivity curve are  $a = 4.35, b = 1.05$ as found by
\cite{christiansen2015}. 
The integral boundary $x = \rm{MES}-4.1-(\rm{MES}_{thresh}-7.1)$.
  MES (Multiple Event Statistics) 
depends on the transit depth, observation errors and the number of 
observed transits:
\begin {equation}
\rm MES=\dfrac{\sqrt{N_{\rm trn}}\Delta} {\sigma_{\rm cdpp}}
\label{equation:mes}
,\end {equation}
where $N_{\rm trn} = (T_{\rm obs}/P)\times f_{\rm duty}$ is the 
expected number of transits, $T_{\rm obs}$ is the time baseline
of observational coverage for a target and $f_{\rm duty}$ is the
observing duty cycle. The $f_{\rm duty}$ is defined as 
the fraction of $T_{\rm obs}$ with valid observations. 
$\Delta$ is the expected transit signal depth. 
The robust root-mean-square (RMS) combined 
differential photometric precision (CDPP) $\sigma_{\rm cdpp}$ 
is an empirical estimate of the noise in the relative flux time 
series observations. The  ${\rm MES_{thres}}$ reflects 
the transit-signal significance level achieved by the 
transiting planet search (TPS) module.
The probability that a transit of a planet will be detected 
during the time of observation is
\begin {equation}
P_{\rm det}(R,P)= P_K(R,P)P_{\rm win}
\label{equation:detect1}
,\end {equation}
where $P_{\rm win}$ is the window function probability of detecting 
at least three transits. It can be explicitly written out in 
the binomial approximation as
\begin {equation}
\begin{gathered}
P_{\rm win}=1-(1-f_{\rm duty})^M-Mf_{\rm duty}(1-f_{\rm duty})^{(M-1)}\\
-0.5M(M-1)f_{\rm duty}^2(1-f_{\rm duty})^{M-2}
\end{gathered}
\label{equation:window}
,\end {equation}
where $M = T_{\rm obs}/P_{\rm orb}$.
Using these equations and parametric model for 
the planet distribution function (PLDF) presented by 
\cite{youdin2011}
we can estimate the expected number of detectable exoplanets in each
spatial bin, $N_{\rm exp}$:
\begin {equation}
N_{\rm exp}=F_{\rm 0}C_{\rm n}\int_{P_{\rm min}}^{P_{\rm max}}\int_{R_{\rm min}}^{R_{\rm max}}
\left[\sum\limits_{j=1}^N\eta_j(R,P)\right]\times g(R,P)dPdR ,
\label{equation:nexppl}
\end {equation}  
where, $g(R,P)$ describes the exoplanet distribution by radii and orbital 
periods:

\begin {equation}
\begin{gathered}
g(R,P)=\\
\begin{cases}
   \left(\frac{P}{P_0}\right)^{\beta_1}\left(\frac {R}{R_0}\right)^{\alpha_1},\: \rm{if}\: R<R_{\rm brk}\: \rm {and}\: P<P_{\rm brk}
   \\
   \left(\frac{P}{P_0}\right)^{\beta_2} \left(\frac{P_{\rm brk}}{P_0}\right)^{\beta_1-\beta_2}    
   \left(\frac {R}{R_0}\right)^{\alpha_1}, \: \rm{if}\: R<R_{\rm brk}\: \rm {and}\: P\ge P_{\rm brk}
   \\
   \left(\frac{P}{P_0}\right)^{\beta_1}\left(\frac {R}{R_0}\right)^{\alpha_2}  \left(\frac{R_{\rm brk}}{R_0}\right)^{\alpha_1-\alpha_2}
    ,\: \rm{if}\: R\ge R_{\rm brk}\: \rm {and}\: P<P_{\rm brk}
   \\
   \left(\frac{P}{P_0}\right)^{\beta_2} \left(\frac{P_{\rm brk}}{P_0}\right)^{\beta_1-\beta_2}    
   \left(\frac {R}{R_0}\right)^{\alpha_2}\left(\frac{R_{\rm brk}}{R_0}\right)^{\alpha_1-\alpha_2}
   \: \rm{if}\: R\ge R_{\rm brk}\: \rm {and}\: P\ge P_{\rm brk.}
 \end{cases}
 \end{gathered}
\label{equation:nexpt}
\end {equation} 
This PLDF model is modified by the per-star transit survey 
effectiveness (or perstar pipeline completeness), $\eta_j$, 
summed over $N$ targets in the sample. It can be expressed
as $\eta_j = P_{\rm {det},j}\times P_{\rm {tr},j}$ where
$P_{\rm{tr},j}=(R_{\star}/a)(1-e^2)$ is 
the geometric probability of a planet to transit the star.  
The $C_n$ is determined from the normalisation requirement,
\begin {equation}
\int_{P_{\rm max}}^{P_{\rm min}}\int_{R_{\rm max}}^{R_{\rm min}}C_n g(R,P)dRdP=1
\label{equation:cnorm}
.\end {equation} 
$F_0$ is an average number of planets per star in the sample, or real 
exoplanet occurrence. We calculate it via a maximisation of 
the Poisson likelihood, $L$, of the data from a survey that 
detects $N_{\rm pl}$ planets (in our case, it is the number 
of detected exoplanets in each bin) around $N$ survey targets:
\begin {equation}
L \sim \left[ F_0^{N_{\rm pl}}C_n^{N_{\rm pl}}\prod\limits_{i = 1}^{N_{\rm {pl}}}g(R_i,P_i)
\right]exp\left(-N_{\rm exp}\right)
\label{equation:maxlikely}
.\end {equation} 
As it follows from maximisation condition, in the case of the fixed 
parameters for $R_{\rm brk}$, $P_{brk}$, $\alpha_1$, $\alpha_2$,
$\beta_1$, $\beta_2,$
the value of exoplanet occurrence for each bin may be obtained 
as a point where  derivative of $L$ is equal to zero:
\begin{equation}
\frac{\partial L}{\partial F} = 0
\label{equation:deriv}
.\end {equation}
Upon solving this equation, we get an exoplanet occurrence:

\begin {equation}
F_0=\frac{N_{\rm pl}}{C_{\rm n}\int_{P_{\rm min}}^{P_{\rm max}}\int_{R_{\rm min}}^{R_{\rm max}}
\left[\sum\limits_{j=1}^N\eta_j(R,P)\right]\times g(R,P)dPdR}
\label{equation:findf}
.\end {equation} 
 Rather than simultaneously determine all 
parameters in the model, we reduce the dimensionality 
of the problem and impose a prior on the model by 
assuming that the PLDF parameters are fixed at 
values determined in the literature.  This allows us to
study the real planet occurrence in more detail,
rather than focusing on a global fit allowing for 
all parameters to vary.We use the following values of free 
parameters:
$R_{\rm brk}=0.94,\,\alpha_1=19.68,\,\alpha_2=-1.78,\,\beta_2=-0.65$
from \citep{burke15}. Unfortunately, the authors only analysed a 
region from $50 < P < 300$ days and  $0.75 < R < 2.5~R_{\rm Earth}$, 
so we adopted $P_{\rm brk}=7.0$ and $\beta_1=2.23$ from 
\citep{youdin2011}. In this work, the authors use the same sample of 
exoplanets and their values of parameters are in good agreement
with those ones from \citep{burke15} for $P > 50$ days.  
We limit our integrals for determining $N_{exp}$ with 
$P_{\rm min}=0$ d,\,$P_{\rm max}=300$ d. We adopt ranges of 
our subsamples described in previous section as limits for $R$.
We chose 0.5 days
as the bin size for period and $0.02~R_{\rm Earth}$ as the bin size 
for planet radii to be sure that we have enough bins for 
reliable calculation of integrals. 
 During the computation of gradients, we ignore bins where 
the number of expected planets $N_{\rm exp}$ is less 
than one and, simultaneously, the number of detected planets is zero.
The new gradients corresponding to these corrected exoplanet occurrences
in different coordinate systems,
$F_0(r,\alpha,\delta), F_0(r_g,z),F_0(x,y,z)$,
are listed in the Tables~\ref{tab:table2C}, ~\ref{table3C}, and
 ~\ref{tab:table12C}. 
 The real exoplanet frequency is much larger than
the uncorrected frequency of occurrence and the same is true for its gradient error.
As a result, the gradients are significantly larger than 
in the previous case.
However, most of the gradients which were previously found to be significant
became statistically insignificant after the bias correction.
Only the negative gradient along the distance in spherical coordinates, 
$k_r$, is still significant at the $3 \sigma$ level.
It remains to be verified by future observations if this behaviour
is real. Apart from that, the exoplanet frequency of occurrence obtained 
with this method is compatible with a homogeneous space distribution.

 \begin{table*}
\caption{\label{tab:table2C}
Gradients of corrected exoplanet frequency for stars 
up to 16th magnitude in spherical equatorial coordinates 
$(r,\alpha,\delta)$.The one statistically significant gradient (>3$\sigma$)
is highlighted in boldface.
}
\begin{center}
\begin{tabular}{|c|c|c|c|c|c|c|c|c|}
\hline
Sp. type &$N_{stars}$&$N_{planets}$&$k_r,pc^{-1}$&Error&$k_{\alpha},deg^{-1}$&Error&$k_{\delta}, deg^{-1}$ &Error\\
\hline
\multicolumn{9}{|c|}{$R_{planet} \ge 0.75R_{Earth}$} \\
\hline
F& 17191 &370&$\mathbf{ -2.1\times10^{-4}}$&$ 6.7\times10^{-5}$&$   -0.014$&$   7.5\times10^{-3}$&$   -0.013$&$    0.010$\\
G& 26378 &945&$  2.9\times10^{-4}$&$ 1.7\times10^{-4}$&$   -0.016$&$   7.9\times10^{-3}$&$  -2.9\times10^{-3}$&$    0.011$\\
\hline
K& 1919  &82& \multicolumn{6}{|c|}      {Number of objects is too low for satistics }\\ 
\hline
\multicolumn{9}{|c|}{$0.75R_{Earth} \le R_{planet} \le 1.75R_{Earth}$} \\
\hline
F&>>&153&$  -1.1\times10^{-3}$&$ 4.3\times10^{-4}$&$   -0.063$&$    0.033$&$   -0.012$&$    0.043$\\
G&>>&444&$  7.4\times10^{-4}$&$  3.9\times10^{-4}$&$   -0.037$&$    0.01777$&$  5.8\times10^{-3}$&$    0.023$\\
\hline
K&>>&52&\multicolumn{6}{|c|}    {Number of objects is too low for satistics }\\
\hline
\multicolumn{9}{|c|}{$1.75R_{Earth} \le R_{planet} < 3.0R_{Earth}$}\\
\hline
F&>> &126&$ -5.4\times10^{-4}$&$  2.3\times10^{-4}$&$   -0.042$&$    0.021$&$   3.3\times10^{-3}$&$    0.027$\\\
G&>> &332&$  1.6\times10^{-4}$&$  3.0\times10^{-4}$&$   -0.016$&$    0.013$&$   -0.032$&$    0.019$\\
\hline
K &>>&24&\multicolumn{6}{|c|}   {Number of objects is i too low for statistics}\\ 
\hline
\multicolumn{9}{|c|}{ $R_{planet} \ge 3.0R_{Earth}$} \\
\hline
F &>> &91 &$ 1.5\times10^{-5}$&$ 4.6\times10^{-5}$&$   4.1\times10^{-3}$&$   4.7\times10^{-3}$&$  -9.6\times10^{-3}$&$   6.5\times10^{-3}$\\
G &>>   &169&$ 2.3\times10^{-4}$&$ 1.4\times10^{-4}$&$   5.3\times10^{-3}$&$   6.5\times10^{-3}$&$   1.9\times10^{-3}$&$   9.1\times10^{-3}$\\
\hline
K &>>&6&\multicolumn{6}{|c|}    {Number of objects is too low for statistics}\\
\hline
\end{tabular}
\end{center}
\end{table*}

\begin{table*}
\label{table3C}
\caption{Gradients of corrected exoplanet frequency for stars 
up to 16th magnitude in cylindrical coordinates ($r_g,z$).
There are no gradients more significant than 3$\sigma$.}
\begin{center}
\begin{tabular}{|c|c|c|c|c|}
\hline
Sp.t. &$k_{rg}, pc^{-1}$&Error&$k_z,pc^{-1}$&Error\\
\hline
\multicolumn{5}{|c|}{$R_{planet} \ge 0.75R_{Earth}$} \\
\hline
F&$ -1.2\times10^{-4}$&$  5.0\times10^{-4}$&$ -6.3\times10^{-4}$&$  3.7\times10^{-4}$\\
G&$ -5.1\times10^{-4}$&$  8.7\times10^{-4}$&$  9.9\times10^{-4}$&$  7.2\times10^{-4}$\\
\hline
K&\multicolumn{4}{|c|}  {Number of objects is too low for statistics}\\
\hline
\multicolumn{5}{|c|}{$0.75R_{Earth} \le R_{planet} < 1.75R_{Earth}$} \\
\hline
F&$  1.4\times10^{-3}$&$   3.0\times10^{-3}$&$  -2.4\times10^{-3}$&$   2.0\times10^{-3}$\\
G&$ -3.6\times10^{-4}$&$   1.9\times10^{-3}$&$   3.2\times10^{-3}$&$   1.6\times10^{-3}$\\
\hline
K &\multicolumn{4}{|c|} {Number of objects is too low for statistics}\\
\hline
\multicolumn{5}{|c|}{$1.75R_{Earth} \le R_{planet} < 3.0R_{Earth}$} \\
\hline
F&$   2.4\times10^{-3}$&$   1.4\times10^{-3}$&$ -2.1\times10^{-4}$&$  1.1\times10^{-3}$\\
G&$  -1.8\times10^{-3}$&$   1.5\times10^{-3}$&$ -2.4\times10^{-4}$&$  1.2\times10^{-3}$\\
\hline
K &\multicolumn{4}{|c|} {Number of objects is too low for statistics}\\
\hline
\multicolumn{5}{|c|}{ $R_{planet} \ge 3.0R_{Earth}$} \\
\hline
F&$ -5.2\times10^{-4}$&$  3.3\times10^{-4}$&$ -2.2\times10^{-4}$&$ 2.3\times10^{-4}$\\
G&$ -6.7\times10^{-5}$&$  7.2\times10^{-4}$&$  4.8\times10^{-4}$&$ 5.9\times10^{-4}$\\
\hline
K &\multicolumn{4}{|c|} {Number of objects is too low for statistics}\\
\hline
\end{tabular}
\end{center}
\end{table*}

\begin{table*}
\caption{\label{tab:table12C}
Gradients of corrected exoplanet frequency for stars 
up to 16th magnitude in Cartesian coordinates $(x, y, z)$.
There are no gradients more significant than 3$\sigma$. }
\begin{center}
\begin{tabular}{|c|c|c|c|c|c|c|}
\hline
Sp. type &$k_x,pc^{-1}$&Error&$k_y,pc^{-1}$&Error&$k_z, pc^{-1}$ &Error\\
\hline
\multicolumn{7}{|c|}{$R_{planet} \ge 0.75R_{Earth}$} \\
\hline
F&$  6.0\times10^{-4}$&$ 5.1\times10^{-4}$&$ -5.5\times10^{-4}$&$  1.9\times10^{-4}$&$  7.7\times10^{-4}$&$  5.7\times10^{-4}$\\
G&$  5.7\times10^{-4}$&$ 9.8\times10^{-4}$&$ -1.0\times10^{-4}$&$  3.7\times10^{-4}$&$  1.1\times10^{-3}$&$  1.0\times10^{-3}$\\
\hline
K&\multicolumn{6}{|c|}  {Number of objects is too low for statistics}\\
\hline
\multicolumn{7}{|c|}{$0.75R_{Earth} \le R_{planet} \le 1.75R_{Earth}$} \\
\hline
F&$   2.9\times10^{-3}$&$   3.3\times10^{-3}$&$  -3.0\times10^{-3}$&$   1.1\times10^{-3}$&$ 4.5\times10^{-3}$&$   3.1\times10^{-3}$\\
G&$   6.0\times10^{-4}$&$   2.1\times10^{-3}$&$  -2.4\times10^{-4}$&$   8.0\times10^{-4}$&$ 3.5\times10^{-3}$&$   2.1\times10^{-3}$\\
\hline
K&\multicolumn{6}{|c|}  {Number of objects is too low for statistics}\\
\hline
\multicolumn{7}{|c|}{$1.75R_{Earth} \le R_{planet} < 3.0R_{Earth}$}\\
\hline
F&$   4.4\times10^{-5}$&$  1.5\times10^{-3}$&$ -1.4\times10^{-3}$&$   5.8\times10^{-4}$&$   3.6\times10^{-3}$&$   1.8\times10^{-3}$\\
G&$   2.2\times10^{-3}$&$  1.7\times10^{-3}$&$ -4.3\times10^{-4}$&$   6.3\times10^{-4}$&$   2.7\times10^{-4}$&$   1.7\times10^{-3}$\\
\hline
K &\multicolumn{6}{|c|} {Number of objects is too low for statistics}\\ 
\hline
\multicolumn{7}{|c|}{ $R_{planet} \ge 3.0R_{Earth}$} \\
\hline
F&$ 3.2\times10^{-4}$&$  3.7\times10^{-4}$&$ 1.1\times10^{-5}$&$  1.3\times10^{-4}$&$ -2.9\times10^{-4}$&$  3.7\times10^{-4}$\\
G&$ -4.0\times10^{-4}$&$ 7.9\times10^{-4}$&$ 4.1\times10^{-4}$&$  3.0\times10^{-4}$&$ -3.1\times10^{-4}$&$  8.1\times10^{-4}$\\
\hline
K &\multicolumn{6}{|c|} {Number of objects is too low for statistics}\\
\hline
\end{tabular}
\end{center}
\end{table*}

\section{ Spatial gradients of the metallicity}

The above-mentioned drop in the uncorrected exoplanet frequency 
with the height above
the Galactic plane might be related to the decrease of metallicity with 
the Galactocentric radius and height above the galactic plane 
\citep{bergemann14,duong18,esteban17,esteban18}.
To check whether the above-mentioned $k_r, k_z, k_y$ gradients
may be related to the metallicity behaviour, we also calculated 
the gradients of metallicity in the Kepler field exploiting 
the [Fe/H] values of all Kepler target stars. 
We calculated average metallicity in each of 105 spatial bins
applying the same restrictions for the spectral types and volume as
for the 16mag limited sample before.
Metallicity was approximated by the similar linear functions: 
\begin {equation}
[Fe/H](r,\alpha, \delta)=k_r r+k_{\alpha} \alpha+k_{\delta} \delta+k_0.
\label{equation:linear1}
\end {equation} 
\begin {equation}
[Fe/H](r_g, z)=k_{rg} r_g+ k_z z+ k_0 .
\label{equation:lineargrad}
\end {equation} 
\begin {equation}
[Fe/H](x, y, z)=k_{x} x+ k_y y+k_z z+ k_0 .
\label{equation:linearg1}
\end {equation} 
Coefficients of the fit are listed in the Tables ~\ref{tab:table4},
 \ref{tab:table5}, and ~\ref{tab:table13b}.

In the equatorial coordinates $(r,\alpha,\delta)$, the gradients along 
the right ascension or declination are not statistically significant.
The gradients in the distance are significant at the $2\sigma$ level
but they are so low that they represent changes in $[Fe/H]$ 
at the 0.01 dex level which can hardly have any effect on the planet
formation and may be just an artifact of the method used.

Gradients in the Galactic ($r_g, z$) and Cartesian ($x, y, z$)
 coordinates are more interesting. For G stars we observe a 
significant (>3$\sigma$) increase of metallicity with the 
Galactocentric radius. This gradient represents
only about 0.06 dex increase of metallicity across our volume. 
It is not in agreement with the above-mentioned previous 
studies, which report the opposite tendency. 
For F stars in Cartesian coordinates, we observe statistically 
significant positive gradient $k_y$ and negative gradient $k_z$
 (which is consistent with previous studies). It is worth noting
that the range of $z$ values is comparable with the thickness of the 
Galactic disc, while the range of $y$ values covers only a minor part 
of Galaxy. Thus, the negative gradient of metallicity might have
slightly contributed to the decrease of the uncorrected exoplanet 
occurrence with the $z$ coordinate. 
Nevertheless, there are no statistically significant metallicity
gradients which could explain observed trends in uncorrected 
exoplanet occurrence with distance and $y$ coordinate. 
 It is also useful to note that the metallicities from 
the Kepler catalog were found rather inaccurate at this level
\citep{petigura2017}

\section{Relative exoplanet frequency and metallicity in the vicinity
of open clusters} 
\label{sec5}

There are four open stellar clusters which belong to the Kepler field of 
view, listed in the catalogues of \cite {2002A&A...389..871D} and
\cite{cantat2018}: 
NGC 6811, NGC 6819, NGC 6866, and NGC 6791 
(Fig. ~\ref {figure:planets}). 
Apart from these, there is also one cluster, Skiff J1942+38.6, located on the very
edge of the Kepler field.
The main characteristics of 
these clusters are presented in Table \ref{tab:table7}.
Here the age and [Fe/H] are taken from \cite {2002A&A...389..871D}
and coordinates, distance, and radii are from $Gaia$ DR2
\citep{cantat2018}.
Most of them are about a billion years old.
Unfortunately, the location of Skiff J1942 is just beyond the edge of the
Kepler field. That is why it was excluded from this study.

To check if the presence of a cluster  influences the exoplanet 
occurrence, we estimated the exoplanet frequency at different distances 
from the cluster spatial centre.
We note that the radii of the clusters (see Table \ref{tab:table7}) are 
significantly smaller
than the typical $1\sigma$ precision of the $Gaia$ DR2 at the distance
of the cluster.
That is why we decided to define a cylinder-like volume around each 
cluster.
The half-length of the cylinder, $l_1$, is equal to $2 \sigma$ 
precision 
of the $Gaia$ distance measurements. The radius of the cylinder, $r_1$, 
is arbitrarily set to 20 pc. Smaller radii would limit the volume 
heavily and one would not find many planets for the statistics. 
Larger radii would cause the angular radius of the cylinder to be 
larger than the field of individual Kepler chips.
Then we calculate the planet frequency within this cylinder
and compare it with the planet frequency in an outer shell of 
the cylinder. 
The outer shell will have the shape of a hollow
cylinder which is two times bigger extending from 
$r_1$ to $r_2=2\times r_1$ and from $l_1$ to $l_2=2\times l_1$.
The results are summarised in  Table \ref{tab:table8}.
\begin{table*}
\caption{\label{tab:table8}Exoplanet frequency and metallicity in 
the vicinity of open clusters. Close vicinity of the cluster
is represented by the inner shell/cylinder with the size $r_1,l_1$
in pc.
It is compared to a more distant outer shell which is two times bigger.
$N_{st1}, N_{st2}, N_{pl1}, N_{pl2}$, are the number of Kepler target 
stars and exoplanet candidates within the inner and outer shells, 
respectively.
$f_{pl1}, f_{pl2}$ are the relative exoplanet frequencies in the inner
and outer shells, respectively.
$[Fe/H]_{1}, [Fe/H]_{2}$ are the metallicities of the inner and outer 
shells, respectively.}
\begin{center}
\begin{tabular}{|c|c|c|c|c|c|c|c|c|c|c|}
\hline
Name &  $r_1$ & $l_1$ & $N_{st1}$& $N_{pl1}$&$f_{pl1}$& $[Fe/H]_1$&$N_{st2}$&$N_{pl2}$&$f_{pl2}$&$[Fe/H]_2$\\
\hline
NGC 6811 & $20$ & $60$ & $656$ & $22$ & $0.034\pm 0.007$ & $-0.127\pm 0.009$ & $1890 $& $52$ &$ 0.028\pm 0.004$ &$-0.151\pm 0.005$\\
NGC 6819 & $20$ &$200$ & $57$  & $0$  &$ 0 $             &$ -0.505\pm 0.093$ &$ 226  $& $ 2$ &$0.009\pm 0.006$  &$-0.273\pm 0.047$ \\
NGC 6866 & $20$ &$100$ & $240$ & $5$  &$ 0.021\pm 0.009$ & $-0.167\pm 0.015$ & $402$  & $9 $ &$ 0.022\pm 0.004 $&$-0.192\pm 0.011$\\
\hline
\end{tabular}
\end{center}
\end{table*}

There are too few planets in the vicinity of NGC 6819 to draw any 
conclusions. The exoplanet frequency in the inner and outer
cylinders around NGC 6811 and NGC 6866 indicate no variation within 
current error bars, which is in agreement with the observations 
\citep{meibom13,kurtis2018}
 and theoretical predictions \citep{2001MNRAS.322..859B,fuji2018}.
We assumed that enhanced crowding in the clusters 
did not impact the planet detection efficiency since we calculated 
the occurrence rate per Kepler target star and not the absolute occurrence 
rate. However, it can be expected that enhanced crowding in
clusters may increase the probability of false positives
(background eclipsing binaries) since the number of 'blending' stars 
increases.

As a byproduct of this analysis, we also studied the behaviour of 
the metallicity in the vicinity of these clusters.
We do not find any significant difference in the metallicity
in the inner and outer cylinders of NGC 6811 and NGC 6866.
However, we find a slightly lower metallicity
in the inner shell of NGC 6819 (-0.505) compared to its outer shell
(-0.273).

\section{Exoplanet candidates in the open clusters} 
\label{sec6}
\subsection{Location and proper motion criteria}
As a byproduct of our analysis of exoplanet occurrence in the 
vicinity of open clusters, we also searched for new exoplanet 
candidates that are members of these open clusters.
Several groups searched for exoplanets in the clusters located 
in the Kepler field. \cite {2005AJ....129.2856M} searched 
for exoplanets in  NGC 6791 (19h 20m 53,0s; 
+37\textdegree  46' 18'') but did not find any.
In the Kepler data, \cite{meibom13} discovered two 
mini-Neptunes (Kepler-66b and Kepler-67b or 
KIC 9836149b \& KIC 9532052b, respectively) orbiting
 Sun-like stars in the cluster NGC6811. 
The authors argue that such small planets can form and survive
in a dense cluster environment and that it implies that the 
frequency and properties of planets in open clusters are 
consistent with those of planets around field stars 
in the Galaxy.

To establish whether a Kepler host star belongs to an open 
cluster we applied several criteria.
First, we applied a proper motion criterion that assumes that
proper motions of cluster members are distributed 
normally. We calculate for each Kepler host star its proper 
motion
membership probability $P_{\mu}$ as
\begin {equation}
P_{\mu}= exp\left[-\frac{(\mu_{\alpha,cl}-\mu_{\alpha,st})^2}
{2(\sigma_{\alpha,cl}^2+\sigma_{\alpha,st}^2)}\right]
exp\left[-\frac{(\mu_{\delta,cl}-\mu_{\delta,st})^2}
{2(\sigma_{\delta,cl}^2+\sigma_{\delta,st}^2)}\right],
\label{equation:probability1}
\end {equation} 
where $\mu_{\alpha,cl}, \mu_{\delta,cl},$ are the proper motions of 
the cluster in right ascension and declination;
$\sigma_{\alpha,cl}, \sigma_{\delta,cl}$ are the dispersions of 
the proper motion distribution in right ascension and declination
taken from \cite{cantat2018};
$\mu_{\alpha,st}, \mu_{\delta,st}$ are the proper motions of 
individual stars; and
$\sigma_{\alpha,st}, \sigma_{\delta,st}$ are the 
errors in proper motion 
of individual stars, respectively.

Our second criterion is the tangential angular distance from 
the cluster centre. Following \cite{kraus2007}, we use 
an exponential function to describe
this membership probability $P_{\rho}$ which depends on
equatorial coordinates of potential members as
\begin {equation}
P_{\rho}=exp(-\frac{\rho}{\rho_{cl}})
\label{equation:probability2}
,\end {equation}
where $\rho$ is the angular distance from the star to the centre of 
the cluster, and $\rho_{cl}$ is a parameter which characterises 
the angular size of cluster. The angular distance is calculated 
in the following way:
\begin {equation}
\begin{gathered}
\rho=\arccos(\cos(90^{\circ}-\delta_{cl})\cos(90^{\circ}-\delta_{st})+\\
+\sin(90^{\circ}-\delta_{cl})\sin(90^{\circ}-\delta_{st})
\cos(\alpha_{st}-\alpha_{cl})),
\end{gathered}
\label{equation:dist}
\end {equation}
where $(\alpha,\delta)_{cl}$ are coordinates of the cluster centre and
$(\alpha,\delta)_{st}$ are coordinates of an individual star.
To obtain the parameter $\rho_{cl}$ we took a list of cluster members 
from \cite{cantat2018}, built their distribution according 
to the tangential angular distance and fitted for this parameter. 
The resulting values of $\rho_{cl}$ for each cluster are listed in 
the Table \ref{tab:table7}.

Our third criterion is distance.
There is no reason to suspect that the distribution of radial 
distances of cluster members is different from the tangential 
distances. However, the error in radial distance to the stars 
in $Gaia$ DR2 is much larger than the error in the tangential 
distance due to small uncertainties in $\alpha$ and $\delta$. 
For NGC 6811, which is located at the distance 
of 1112 pc, the error in the radial distance is about 30 pc.
It is several times larger than the radius of the cluster. 
That is why we assume that the distribution in radial distances 
is normal and govern fully by the errors of the distance measurements.
Hence, we define its probability $P_r$ as
\begin {equation}
P_{r}=exp\left[-\frac{(r_{cl}-r_{st})^2}{2(\sigma_{r,cl}^2+\sigma_{r,st}^2)}\right]
\label{equation:probability3}
,\end {equation}
where $r_{cl}$ is the mean distance from the Sun to the cluster, 
$\sigma_{r,cl}$ is its dispersion (all listed in Table \ref{tab:table7}),
$r_{st}$ is the distance to an individual star, and
$\sigma_{r,st}$ is its error.

It would also be possible to use another criterion based on radial velocities.
Unfortunately, $Gaia$ DR2 does not contain radial velocities of 
our exoplanet host stars which are members of clusters because 
they are too faint.

Finally, we calculate a total probability $P_{tot}$ of membership 
for individual stars as a product of the three above-mentioned 
probabilities:
\begin {equation}
P_{tot}=P_{\mu}P_{\rho}P_r
\label{equation:probability4}
.\end {equation}

Using this equation, we checked all Kepler exoplanet host stars 
for their membership in the above-mentioned four open clusters.
A few top-ranked stars with the highest $P_{tot}$ 
values are listed in Table \ref{tab:table9}.
Consequently, we selected four exoplanet host star candidates that 
are the most probable cluster members of NGC 6811 
(Kepler-66, Kepler-67, KIC 9655005, KIC 9533489)
and one less promising member of this cluster (KIC 9776794). 
The first two of these have already been mentioned
in \cite {meibom13}. Our two members have even
higher cluster probabilities, but they also have very high false 
positive probabilities, indicating that they may not be exoplanets.
In NGC 6866, we found two highly probable members which are also
exoplanet host star candidates (KIC 8331612, KIC 8396288).
We found no good candidates in other clusters.

\subsection{Colour-magnitude and colour-period diagrams}

\begin{figure}
\center{\includegraphics[width=\linewidth]{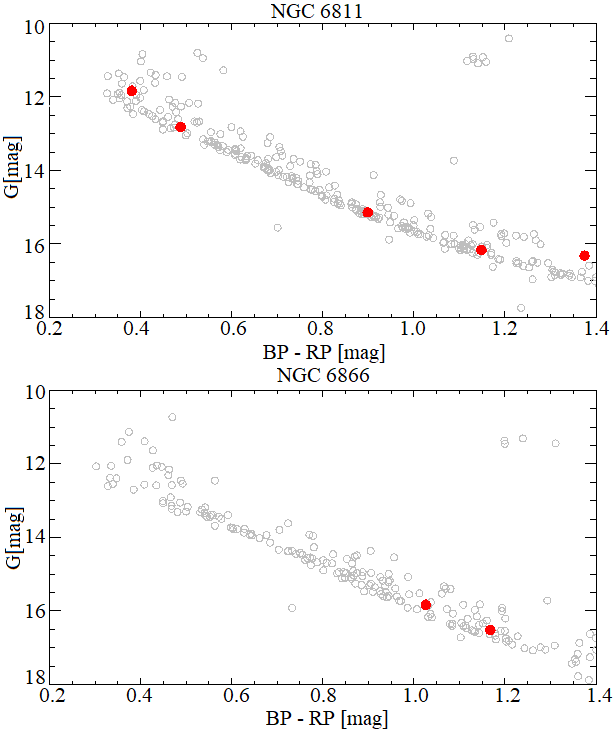}}
\caption{Colour-magnitude diagram of clusters NGC 6811(upper panel) 
and NGC 6866 (lower panel). Exoplanet host stars are marked with red 
circles. }
\label{figure:color_magnitude}
\end{figure}

To verify the cluster membership of these top seven exoplanet candidates,
we place them in a colour-magnitude diagram 
(Fig. \ref{figure:color_magnitude})
together with other cluster members taken from \cite{cantat2018}.
All of them fit among the other members very well.
Only KIC 9776794, which is the least likely member out of seven, lies
 slightly above the main sequence.
We do not use this diagram as a separate criterion 
(only as a verification) since this information was already
taken into account in the above-mentioned radial distance 
criterion.
The cluster membership can be verified also by the rotational periods 
of the stars. Previous studies have shown that cluster members form 
a relatively narrow and well-defined sequence in the colour-period 
diagram (CPD). For example, FGK dwarfs in the Hyades 
\citep{radick1987,delorme2011,douglas2018}, 
Praesepe \citep{delorme2011, Kovacs2014}, 
Coma Berenices \citep{cameron2009}, Pleiades \citep{stauffer2016}, 
and M 37 \citep{hartman2009}. \cite{barnes2015} constructed a CPD 
for M48 and derived its rotational age using gyro-chronology. 
This sequence in the CPD is followed closely by a theoretical 
isochrone and looks similar to the CPD of other open clusters 
of the same age. \cite{barnes2016} found that rotation periods of 
M67 members delineate a sequence in the CPD reminiscent of that 
discovered first in the Hyades cluster. Similar sequence we can see in 
the mass-period diagram (which is equivalent to CPD) of NGC 752 
\citep{agueros2018}. In general, the rotational evolution 
theory is described in \cite{barnes2010}, \cite{saders2019}.
\cite{Kovacs2014} also found that exoplanet host stars may have
shorter periods than predicted due to the tidal interaction and angular 
moment exchange between star and planet, so host star can lie below 
the sequence of other members on CPDs.

\begin{figure}
\center{\includegraphics[width=\linewidth]{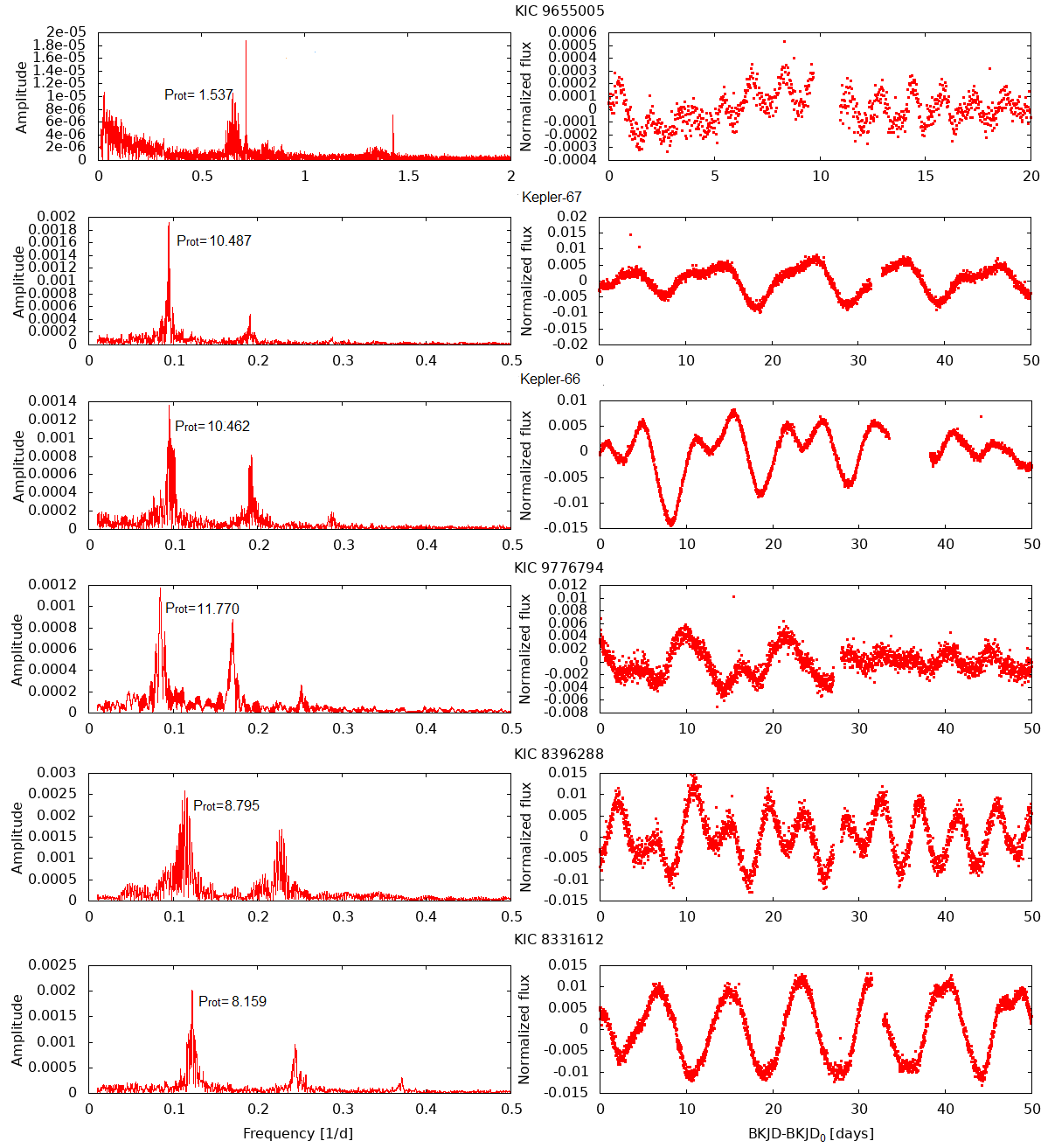}}
\caption{Kepler long cadence light curves (right panel) and
 their power spectra (left panel). }
\label{figure:ROTATION}
\end{figure}

For this purpose, we determined the rotational periods 
of our top seven candidates.
We used the Kepler long cadence data in the form of PDCSAP flux 
as a function of a Kepler barycentric Julian day (BKJD), which is a Julian day minus 2454833.
The periods were searched with the Fourier method \citep{deeming1975}. 
We estimated errors in rotation periods with the Monte-Carlo method.
The results are listed in Table \ref{tab:table9}. 
The light curves and power spectra are shown in 
Fig. \ref{figure:ROTATION}.
One of the stars, KIC 9533489, had already been studied by \cite{bognar2015}
who discovered that it is a $\gamma$ Dor/$\delta$ Scu-type pulsator
and identified a few probable rotational periods.

\begin{figure}
\center{\includegraphics[width=\linewidth]{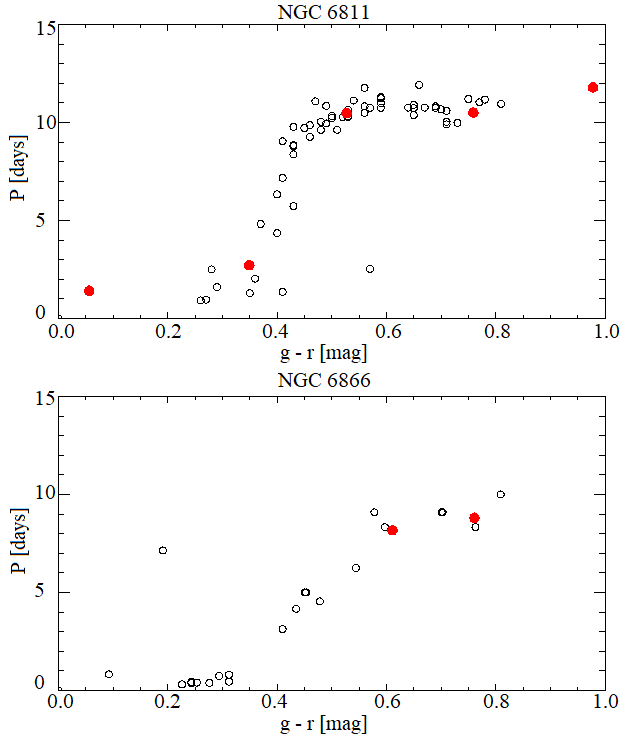}}
\caption{Colour-period diagram of clusters NGC 6811(upper panel) 
and NGC 6866 (lower panel). Exoplanet host stars are marked with red 
circles. }
\label{figure:color_period}
\end{figure}

Next, we constructed colour-period diagrams for both clusters.
They are shown in Fig. \ref{figure:color_period},
together with other stars that are members of those clusters.
The rotational periods of the other members of NGC 6811 and NGC 6819 
were taken from \cite{meibom2011} and \cite{balona2013}, 
respectively.

The location of KIC 9533489, Kepler-66, and Kepler-67 is in
very good agreement with the location of other cluster members of 
NGC 6811. KIC 9655005 and KIC 9776794 are beyond the range of
comparison stars since they are too blue or red, but they also seem
to fit well into the pattern.
The location of both exoplanet host stars from the NGC 6866 
(KIC 8331612, KIC 8396288) are also in very good agreement 
with the location of other cluster members.
This adds credibility to the cluster membership of these exoplanet 
host stars.

\section*{ Conclusions}

Our main findings are summarised below.
\begin {itemize}
\item 
We searched for inhomogeneities in the frequency of exoplanet occurrence
on the scales of hundreds of parsecs. Data from $Gaia$ and Kepler
satellites were used for this purpose.
We found statistically significant gradients of the  uncorrected
exoplanet frequency along the distance, 
Galactic longitude $l=90^{\circ}$, and height above the Galactic plane.
We argue that these gradients are most probably caused by a single
observational bias of undetected small planets around faint stars.
When we corrected for this bias the gradients 
became statistically insignificant.
 Only the gradient of planet occurrence with distance for F stars
remains significant at the $3 \sigma$ level.
We did not find any other significant gradients in the Cartesian,
Galactocentric nor spherical coordinate systems.
Consequently, apart from that one gradient, the spatial distribution of 
exoplanets in the Kepler field of view is compatible with a homogeneous 
one.

\item
We searched for the inhomogeneities in the exoplanet frequency
on the scales of tens of parsecs in the vicinity of open clusters
based on Kepler and $Gaia$ data.
We do not find a significant difference in the exoplanet occurrence
in the vicinity of the clusters.

\item 
The metallicity of our G star sample was found to 
increase with the Galactocentric radius and slightly decrease
with the distance (at 2$\sigma$ level).
However, the metallicity of the F star sample increases slightly
with the distance and $y$ coordinate and drops 
with the Galactocentric radius and height above the Galactic plane.
It means that it cannot be the reason for the drop of the exoplanet
frequency around F stars with the distance since the exoplanet 
occurrence increases with the metallicity of the host star.
However, this might have contributed to a drop of  uncorrected
exoplanet frequency around F stars with the height above 
the Galactic plane.
 
\item
We discovered four exoplanet host star candidates which are
members of the open cluster NGC 6811 
(KIC 9655005, KIC 9533489, Kepler-66, Kepler-67).
The last two had already been mentioned
by \cite {meibom13}. 

\item
We found two other promising exoplanet host star candidates 
which belong to the open cluster NGC 6866 (KIC 8396288, KIC 8331612).
All these targets deserve further follow up using spectroscopy.
\end {itemize}
In the future, it might also be interesting to verify and expand this study
using K2 \citep{K2}, TESS\citep{TESS}, or PLATO \citep{PLATO} data in combination with the final $Gaia$ data 
release. 

\section*{ Acknowledgement}
 We thank the anonymous referee for a very careful reading and 
many relevant suggestions.
We are also grateful to Prof. Geza Kovacs, Prof. Nicolas Lodieu, and 
Jaroslav Merc for reading and useful comments on this manuscript.
Our work was supported by the VEGA 2/0031/18 and Erasmus+ 
'Per aspera ad astra simul' projects.
JB was also supported by the Slovak Research and Development Agency
under the contract No. APVV-15-0458.
This work has made use of data from the European Space Agency (ESA) 
mission {\it $Gaia$} (\url{https://www.cosmos.esa.int/gaia}), 
processed by the {\it $Gaia$} Data Processing and Analysis Consortium 
(DPAC, \url{https://www.cosmos.esa.int/web/gaia/dpac/consortium}). 
Funding for the DPAC has been provided by national institutions, 
in particular, the institutions participating in the {\it $Gaia$} 
Multilateral Agreement.

\bibliographystyle{aa} 
\bibliography{budaj,my}{}

\clearpage
\begin{appendix}

\section{Spatial gradients of uncorrected exoplanet frequency}
\begin{table*}[b]
\caption{\label{tab:table2}
Same as in the Table \ref{tab:table2C}
but the gradients of exoplanet frequency are not corrected for 
observational biases. Statistically significant gradients (>3$\sigma$)
are highlighted in boldface.}
\begin{center}
\begin{tabular}{|c|c|c|c|c|c|c|c|c|}
\hline
Sp. type &$N_{stars}$&$N_{planets}$&$k_r,pc^{-1}$&Error&$k_{\alpha},deg^{-1}$&Error&$k_{\delta}, deg^{-1}$ &Error\\
\hline
\multicolumn{9}{|c|}{$R_{planet} \ge 0.75R_{Earth}$} \\
\hline
F& 17191 &370 & $\mathbf{-1.92\times 10^{-5}}$ &$2.6\times 10^{-6}$&$-5.6\times10^{-4}$&$3.2\times10^{-4}$&$-4.8\times10^{-4}$ &$4.5\times10^{-4}$\\
G& 26378 &945 & $-1.69\times 10^{-5}$&$9.7 \times10^{-6}$&$-6.4\times10^{-4}$&$5.4\times10^{-4}$&$5.9\times10^{-4} $&$7.6 \times10^{-4}$\\
K& 1919  &82  & $-2.95\times 10^{-4}$&$2.2 \times10^{-4}$&$4.4\times10^{-5}$&$2.9\times10^{-3}$&$-2.4\times10^{-3}$&$4.1\times10^{-3}$\\
\hline
\multicolumn{9}{|c|}{$0.75R_{Earth} \le R_{planet} < 1.75R_{Earth}$} \\
\hline
F&>>& 153 & $\mathbf{-1.15\times 10^{-5}}$ &$1.9\times 10^{-6}$&$-5.3\times10^{-4}$&$2.4\times10^{-4}$&$-1.3\times10^{-5}$ &$3.3\times10^{-4}$\\

G&>>&   444 & $-1.14\times 10^{-5}$&$6.8 \times10^{-6}$&$-4.5\times10^{-4}$&$3.8\times10^{-4}$&$-1.9\times10^{-5} $&$5.3 \times10^{-4}$\\

K&>>&   52 & $-2.9\times 10^{-4}$&$2.0 \times10^{-4}$&$-9.1\times10^{-4}$&$2.9\times10^{-3}$&$-2.7\times10^{-3}$&$3.9\times10^{-3}$\\
\hline
\multicolumn{9}{|c|}{$1.75R_{Earth} \le R_{planet} < 3.0R_{Earth}$}\\
\hline
F&>> &126 & $\mathbf{-6.1\times10^{-6}}$&$1.3\times10^{-6}$&$-1.9\times10^{-4}$&$1.6\times10^{-4}$&$-1.8\times10^{-4}$&$2.2\times10^{-4}$\\
G&>> &332 & $-7.6\times10^{-6}$&$5.5\times10^{-6}$&$-1.4\times10^{-4}$&$3.0\times10^{-4}$&$3.0\times10^{-4} $&$4.3\times10^{-4}$\\
\hline
K &\multicolumn{8}{|c|} {24 planets - too few for statistics}\\ 
\hline
\multicolumn{9}{|c|}{ $R_{planet} \ge 3.0R_{Earth}$} \\
\hline
F &>> &91 & $-1.67\times 10^{-6}$ &$1.13\times 10^{-6}$&$1.6\times10^{-4}$&$1.4\times10^{-4}$&$-2.8\times10^{-4}$ &$1.9\times10^{-4}$\\

G &>>   &169 & $2.12\times 10^{-6}$&$3.33 \times10^{-6}$&$-5.2\times10^{-5}$&$1.8\times10^{-4}$&$3.0\times10^{-4} $&$2.6 \times10^{-4}$\\
\hline
K &\multicolumn{8}{|c|} {6 planets - too few for statistics}\\
\hline
\end{tabular}
\end{center}
\end{table*}

\begin{table*}[b]
\caption{\label{tab:table3}
Same as in the Table \ref{table3C}
but the gradients of exoplanet frequency are not corrected for 
the observational biases.
Statistically significant gradients (>3$\sigma$)
are highlighted in boldface.
}
\begin{center}
\begin{tabular}{|c|c|c|c|c|}
\hline
Sp.t. &$k_{rg}, pc^{-1}$&Error&$k_z,pc^{-1}$&Error\\
\hline
\multicolumn{5}{|c|}{$R_{planet} \ge 0.75R_{Earth}$} \\
\hline
F&$1.44\times10^{-5}$&$2.0\times10^{-5}$&$\mathbf{-5.70\times10^{-5}}$&$1.3\times10^{-5}$\\
G&$4.52\times10^{-5}$&$6.4\times10^{-5}$&$-2.19\times10^{-5}$&$5.2\times10^{-5}$\\
K&$3.31\times10^{-4}$&$1.3\times10^{-3}$&$-7.32\times10^{-4}$&$1.2\times10^{-3}$\\
\hline
\multicolumn{5}{|c|}{$0.75R_{Earth} \le R_{planet} < 1.75R_{Earth}$} \\
\hline
F&$1.03\times10^{-5}$&$1.5\times10^{-5}$&$\mathbf{-3.17\times10^{-5}}$&$9.4\times10^{-6}$\\
G&$2.88\times10^{-5}$&$4.5\times10^{-5}$&$-1.16\times10^{-5}$&$3.7\times10^{-5}$\\
K&$4.54\times10^{-4}$&$1.2\times10^{-3}$&$-6.14\times10^{-4}$&$1.2\times10^{-3}$\\
\hline
\multicolumn{5}{|c|}{$1.75R_{Earth} \le R_{planet} < 3.0R_{Earth}$} \\
\hline
F&$1.68\times10^{-5}$&$1.9\times10^{-5}$&$\mathbf{-4.85\times10^{-5}}$&$1.2\times10^{-5}$\\
G&$3.35\times10^{-5}$&$6.2\times10^{-5}$&$-3.44\times10^{-5}$&$5.0\times10^{-5}$\\
\hline
K &\multicolumn{4}{|c|} {24 planets - too few for statistics}\\
\hline
\multicolumn{5}{|c|}{ $R_{planet} \ge 3.0R_{Earth}$} \\
\hline
F&$-2.35\times10^{-6}$&$8.1\times10^{-6}$&$-7.61\times10^{-6}$&$5.2\times10^{-6}$\\
G&$1.16\times10^{-5}$&$2.2\times10^{-5}$&$1.25\times10^{-5}$&$1.8\times10^{-5}$\\
\hline
K &\multicolumn{4}{|c|} {6 planets - too few for statistics}\\
\hline
\end{tabular}
\end{center}
\end{table*}
\begin{table*}[b]
\caption{\label{tab:table12}
Same as in the Table \ref{tab:table12C}
but the gradients of exoplanet frequency are not corrected for 
the observational biases.
Statistically significant gradients (>3$\sigma$)
are highlighted in boldface.
}
\begin{center}
\begin{tabular}{|c|c|c|c|c|c|c|}
\hline
Sp. type &$k_x,pc^{-1}$&Error&$k_y,pc^{-1}$&Error&$k_z, pc^{-1}$ &Error\\
\hline
\multicolumn{7}{|c|}{$R_{planet} \ge 0.75R_{Earth}$} \\
\hline
F&$ 1.09\times10^{-5}$&$ 2.0\times10^{-5}$&$\mathbf{-2.57\times10^{-5}}$&$ 7.3\times10^{-6}$&$ 1.08\times10^{-5}$&$ 2.1\times10^{-5}$\\ 
G&$-3.82\times10^{-6}$&$ 7.4\times10^{-5}$&$-2.57\times10^{-5}$&$ 2.7\times10^{-5}$&$ 3.50\times10^{-5}$&$ 7.5\times10^{-5}$\\ 
K&$ 3.08\times10^{-4}$&$ 1.6\times10^{-3}$&$-3.66\times10^{-4}$&$ 5.7\times10^{-4}$&$-7.29\times10^{-5}$&$ 1.6\times10^{-3}$\\
\hline
\multicolumn{7}{|c|}{$0.75R_{Earth} \le R_{planet} \le 1.75R_{Earth}$} \\
\hline
F&$ 9.24\times10^{-6}$&$ 1.5\times10^{-5}$&$\mathbf{-1.73\times10^{-5}}$&$ 5.4\times10^{-6}$&$ 1.17\times10^{-5}$&$ 1.5\times10^{-5}$\\
G&$ 3.73\times10^{-6}$&$ 5.2\times10^{-5}$&$-2.32\times10^{-5}$&$ 1.9\times10^{-5}$&$ 4.13\times10^{-5}$&$ 5.3\times10^{-5}$\\
K&$ 2.08\times10^{-4}$&$ 1.5\times10^{-3}$&$-3.80\times10^{-4}$&$ 5.5\times10^{-4}$&$ 7.78\times10^{-5}$&$ 1.5\times10^{-3}$\\
\hline
\multicolumn{7}{|c|}{$1.75R_{Earth} \le R_{planet} < 3.0R_{Earth}$}\\
\hline
F&$ 7.42\times10^{-7}$&$ 1.0\times10^{-5}$&$-7.82\times10^{-6}$&$ 3.7\times10^{-6}$&$ 4.82\times10^{-6}$&$ 1.0\times10^{-5}$\\
G&$ 1.43\times10^{-5}$&$ 4.2\times10^{-5}$&$-1.12\times10^{-5}$&$ 1.5\times10^{-5}$&$-1.35\times10^{-6}$&$ 4.3\times10^{-5}$\\
\hline
K &\multicolumn{6}{|c|} {24 planets - too few for statistics}\\ 
\hline
\multicolumn{7}{|c|}{ $R_{planet} \ge 3.0R_{Earth}$} \\
\hline
F&$ 9.34\times10^{-7}$&$ 8.7\times10^{-6}$&$-5.84\times10^{-7}$&$ 3.1\times10^{-6}$&$-5.68\times10^{-6}$&$ 8.8\times10^{-6}$\\
G&$-2.18\times10^{-5}$&$ 2.5\times10^{-5}$&$ 8.74\times10^{-6}$&$ 9.1\times10^{-6}$&$-4.91\times10^{-6}$&$ 2.6\times10^{-5}$\\
\hline
K &\multicolumn{6}{|c|} {6 planets - too few for statistics}\\
\hline
\end{tabular}
\end{center}
\end{table*}

\clearpage
\section{Spatial gradients of the metallicity}
\begin{table*}[b]
\caption{\label{tab:table4}Gradients of metallicity based on 
all Kepler target stars (a volume-limited 16 mag sample) 
in spherical equatorial coordinates $(r,\alpha,\delta)$.
There are no gradients more significant than 3$\sigma$.}
\begin{center}
\begin{tabular}{|c|c|c|c|c|c|c|c|}
\hline
Sp. type &$k_{r}, pc^{-1}$&Error&$k_{\alpha},deg^{-1}$&Error&$k_{\delta}, deg^{-1}$ &Error\\
\hline
F& $1.41\times10^{-5} $&$5.2\times10^{-6}  $&$1.23\times10^{-3}$&$6.4\times10^{-4}  $&$-1.33\times10^{-3}$&$8.9\times10^{-4}$\\
G& $-3.61\times10^{-5}$&$1.8\times10^{-5}  $&$-2.99\times10^{-4}$&$1.0\times10^{-3} $&$9.30\times10^{-4} $&$1.4\times10^{-3}$\\
K& $-3.63\times10^{-4}$&$1.7\times10^{-4}  $&$-2.86\times10^{-4}$&$2.2\times10^{-3}  $&$8.74\times10^{-4}$&$2.9\times10^{-3}$\\
\hline
\end{tabular}
\end{center}
\end{table*}

\begin{table*}
\caption{\label{tab:table5}Gradients of metallicity based on Kepler 
target stars (a volume-limited 16 mag sample) in coordinates ($r_g,z$).
The one statistically significant gradient ($>3\sigma$) is highlighted in boldface.}
\begin{center}
\begin{tabular}{|c|c|c|c|c|c|c|c|}
\hline
Sp.t. &$k_{rg}, pc^{-1}$&Error&$k_z,pc^{-1}$&Error\\
\hline
F&$-6.61\times10^{-5}$&$3.8\times10^{-5}$&$-1.40\times10^{-5}$&$2.4\times10^{-5}$\\
G&$\mathbf{3.37\times10^{-4}}$&$1.1\times10^{-4}$&$1.25\times10^{-4}$&$9.3\times10^{-5}$\\
K&$9.74\times10^{-4}$&$8.8\times10^{-4}$&$-5.57\times10^{-4}$&$8.3\times10^{-4}$\\
\hline
\end{tabular}
\end{center}
\end{table*}
\begin{table*}
\caption{\label{tab:table13b}Gradients of metallicity based on Kepler 
target stars (a volume-limited 16 mag sample)  in Cartesian coordinates ($x, y, z$).
Statistically significant gradients ($>3\sigma$) are highlighted in boldface.}
\begin{center}
\begin{tabular}{|c|c|c|c|c|c|c|c|c|c|}
\hline
Sp.t. &$k_{x}, pc^{-1}$&Error&$k_y, pc^{-1}$&Error&$k_z, pc^{-1}$&Error\\
\hline
F&$ 3.93\times10^{-5}$&$ 3.7\times10^{-5}$&$ \mathbf{4.56\times10^{-5}}$&$ 1.3\times10^{-5}$&$ \mathbf{-1.63\times10^{-4}}$&$ 3.8\times10^{-5}$\\
G&$-2.28\times10^{-4}$&$ 1.3\times10^{-4}$&$-3.79\times10^{-5}$&$ 4.8\times10^{-5}$&$  2.22\times10^{-4}$&$ 1.3\times10^{-4}$\\
K&$-8.30\times10^{-4}$&$ 1.1\times10^{-3}$&$-1.10\times10^{-4}$&$ 4.0\times10^{-4}$&$ -2.97\times10^{-4}$&$ 1.1\times10^{-3}$\\
\hline

\end{tabular}
\end{center}
\end{table*}

\clearpage

\section{Exoplanet candidates in the open clusters}
\begin{table*}[b]
\caption{\label{tab:table7}Characteristics of the open clusters in 
the Kepler field of view. $\alpha$ and $\delta$ are J2000 equatorial 
coordinates of the cluster centre in degrees, $r$ is the distance to 
the cluster centre in $pc$, $\mu_{\alpha,\delta}$ is the proper motion 
in $mas\,yr^{-1}$, $\sigma_{\alpha,\delta}$ is the dispersion of the proper 
motion distribution($mas\,yr^{-1}$), $\sigma_r$ is the dispersion of the radial distance 
distribution($pc$), $\rho_{cl}$ is a characteristic scale of the cluster in 
degrees, $\rho_{1/2}$ is the radius which contains half of the cluster 
members in degrees, $\log t$ is the logarithm of the cluster age in yrs, 
$v_r$ is the mean radial velocity of the cluster in $km\,s^{-1}$, and 
$\sigma_v$ is its standard deviation. Coordinates, proper motions,
and $\rho_{1/2}$ are from \cite{cantat2018}, age and metallicity are 
from \cite{2002A&A...389..871D}, and radial velocities are from 
\cite{soubiran2018} and \cite{sartoretti2018}.}
\begin{center}
\begin{tabular}{|c|c|c|c|c|c|}
\hline
Name&N6811&N6819&N6866&N6791\\
\hline
$\alpha$&$294.340$&$295.327$&$300.983$&$290.221$\\
$\delta$&$46.378$&$40.190$&$44.158$&$37.778$\\
r&$1112$&$2599$&$1398$&$4530$\\
$\sigma_r$&$68.5$&$539.7$&$ 87.5$&$\--$\\
$\rho_{1/2}$&$0.190$&$0.095$&$0.104$&$0.068$\\
$\log t$&$8.799$&$9.36$&$8.91$&$9.92$\\
$[Fe/H]$&$-0.02$&$-0.02$&$-0.013$&$+0.42$\\
$\mu_{\alpha}$&$-3.399$&$-2.916$&$-1.365$&$-0.421$\\
$\sigma_{\alpha}$&$0.116$&$0.125$&$0.081$&$0.165$\\
$\mu_{\delta}$&$-8.812$&$-3.856$&$-5.743$&$-2.269$\\
$\sigma_{\delta}$&$0.123$&$0.140$&$0.092$&$0.193$\\
$\rho_{cl}$&$0.33$&$0.14$&$0.18$&$\--$\\
$v_r$&$7.40$&$3.31$&$12.83$&$-45.85$\\
$\sigma_v$&$0.43$&$1.93$&$0.86$&$1.64$\\
\hline
\end{tabular}
\end{center}
\end{table*}

\begin{table*}[b]
\caption{\label{tab:table10}Characteristics of our exoplanet host stars
candidates near NGC 6811 and NGC 5866. 
$\alpha$ is right ascension in degrees,
$\delta$ is declination in degrees - both  from \cite{cantat2018},
$g$ is stellar magnitude in SDSS filter from \cite{brown2011}.}
\begin{center}
\begin{tabular}{|c|c|c|c|}
\hline
Name& $\alpha$ & $\delta$ & $g$\\
\hline
\multicolumn{4}{|c|}{NGC 6811}\\
\hline
$KIC 9655005$& $294.201$ & $47.161$&$11.958$\\
$KIC 9533489$& $294.674$ & $46.365$&$13.249$\\
$Kepler-67$&   $294.153$ & $47.039$&$16.868$\\
$Kepler-66$&   $293.982$ & $47.095$&$15.661$\\
$KIC 9776794$& $294.365$ & $45.682$&$16.956$\\
\hline
\multicolumn{4}{|c|}{NGC 6866}\\
\hline
$KIC 8396288$&$301.055$ &     $44.145$&$17.625$\\
$KIC 8331612$&$301.279$ &     $44.135$&$16.426$\\
\hline
\multicolumn{4}{|c|}{NGC 6819}\\
\hline
$Kepler-1625$& $295.429$  &   $39.855$&$\--$\\
\hline
\end{tabular}
\end{center}
\end{table*}

\begin{table*}[b]
\caption{\label{tab:table9}Characteristics of our exoplanet host star 
candidates near NGC 6811 and NGC 6866. $P_{\mu}, P_{\rho}$ and $P_d$
are membership probabilities (see Sect. 6),
 $\Delta G=G_{bp}-G_{rp}$ is 
the colour based on $Gaia$ filters 
\citep{2016A&A...595A...1G,2018A&A...616A...1G},
$g$ and $r$ are stellar magnitudes in SDSS filters \citep{brown2011}. 
$P_{rot} (days)$ is the rotational period of the star, $\Delta P$ is 
its error estimated with the Monte-Carlo method, $P_{orb}$ is 
the planet orbital period in days, $R_p/R_s$ is the planet to star radius
ratio, and $P_f$ is a false positive probability \citep{morton2015}.
$^{(1)}$ is a $\gamma$ Doradus/$\delta$ Scuti star studied in 
\cite{bognar2015}, $^{(2)}$ are the exoplanet host stars studied in 
\cite{meibom13}}
\begin{center}
\begin{tabular}{|c|c|c|c|c|c|c|c|c|c|c|}
\hline
Name & $P_{\mu}$&$P_{\rho}$&$P_d$&$P_{tot}$&$\Delta G$&$g-r$&$P_{rot}\pm \Delta P$&$P_{orb}$&$R_p/R_s$&$P_f$\\
\hline
\multicolumn{11}{|c|}{NGC 6811}\\
\hline
$KIC 9655005$&$0.65$&$0.72$&$1.00$&$0.47$&$0.38$&$0.06$&$1.537\pm0.008$&$1.399$&$0.006$&$1.0$\\
$KIC 9533489$&$0.97$&$0.36$&$0.80$&$0.28$&$0.49$&$0.35$&$2.70$&$197.146$&$0.342$&$1.0$\\
$Kepler-67^{(2)}$&$0.64$&$0.47$&$0.99$&$0.30$&$1.15$&$0.76$&$10.487\pm0.199$&$15.726$&$0.033$&$1.7\times10^{-4}$\\
$Kepler-66^{(2)}$&$0.92$&$0.30$&$0.82$&$0.23$&$0.90$&$0.53$&$10.462\pm0.053$&$17.816$&$0.031$&$5.7\times10^{-3}$\\
$KIC 9776794$&$2.5\times10^{-4}$&$0.68$&$0.71$&$1.2\times10^{-4}$&$1.38$&$0.98$&$11.770\pm0.031$&$18.222$&$0.071$&$0.85$\\
\hline
\multicolumn{11}{|c|}{NGC 6866}\\
\hline
$KIC 8396288$&$0.85$&      $0.42$ &     $0.99$ &     $ 0.36 $&$1.16$&$0.76$&$8.795\pm0.037$&$8.585$&$0.034$&$2.6\times10^{-2}$\\
$KIC 8331612(a)$&$0.32$&      $0.28$ &     $0.90$ &     $0.082$&$1.03$&$0.61$&$8.159\pm0.039$&$13.835$&$0.033$&$2.7\times10^{-3}$\\
$KIC 8331612(b)$&$0.32$&      $0.28$ &     $0.90$ &     $0.082$&$1.03$&$0.61$&$8.159\pm0.039$&$25.697$&$0.033$&$1.2\times10^{-1}$\\
\hline
\multicolumn{11}{|c|}{NGC 6819}\\
\hline
$Kepler-1625$& $3.0\times10^{-15}$ &       $0.10$ &     $0.77$&       $2.3\times10^{-16}$&  $\--$&$\--$ &$\--$&$287.377$&$0.060$&$7.5\times10^{-3}$\\
\hline 
\end{tabular}
\end{center}
\end{table*}

\end{appendix}

\end{document}